\documentclass[11pt,a4paper,final]{article}
 
\usepackage{graphicx}
\usepackage[breaklinks=true, colorlinks=true,linkcolor=blue,urlcolor=blue,citecolor=blue]{hyperref}
\usepackage{slashbox}
\usepackage{sidecap}
\usepackage{authblk}
\usepackage[cm]{fullpage}
\usepackage{cite}

\newcommand{\Eqn}[1]{Eq.~(\ref{#1})}
\newcommand{\Fig}[1]{Figure~\ref{#1}}
\newcommand{\Ref}[1]{Ref~\cite{#1}}
\sidecaptionvpos{figure}{c}
\newcommand{\Omprime}{\ensuremath{\Omega^{\prime}_{\phi}{}}}
\newcommand{\Omprimem}{\ensuremath{\Omega^{\prime}_{\phi m}{}}}
\newcommand{\qprime}{\ensuremath{q^\prime{}}}
\newcommand{\shat}{\ensuremath{\hat{s}}}

\begin{document}

\title{Using a local gyrokinetic code to study global ITG modes in tokamaks.}

\author[1,2]{P. A. Abdoul\thanks{paaa500@york.ac.uk}}
\author[2]{D. Dickinson}
\author[3]{C. M. Roach}
\author[2]{H. R. Wilson}
\affil[1]{Sulaimani University, Faculty of Science, Department of Physics, Al-Sulaimaniyah, Kurdistan Region, Iraq}
\affil[2]{York Plasma Institute, Department of Physics, University of York, Heslington, York, YO10 5DD, UK}
\affil[3]{CCFE, Culham Science Centre, Abingdon, Oxfordshire, OX14 3DB, UK}
\renewcommand\Authands{ and }
\maketitle

\begin{abstract}
\label{abstract}
In this paper the global eigenmode structures of linear ion-temperature-gradient
(ITG) modes in tokamak plasmas are obtained using a novel technique which combines results from the
local gyrokinetic code GS2 with analytical theory to reconstruct global
properties. Local gyrokinetic calculations, using GS2, are performed for a range
of radial flux surfaces, $x$, and ballooning phase angles, $p$, to map out the local
complex mode frequency, $\Omega_{0}(x,p)=\omega_{0}(x,p)+i\gamma_{0}(x,p)$ for a
single toroidal mode number, $n$. Taylor expanding $\Omega_{0}$ about a
reference surface at $x=0$, and employing the Fourier-ballooning representation
leads to a second order ODE for the amplitude envelope, $A\left(p\right)$, which
describes how the local results are combined to form the global mode. The
equilibrium profiles impact on the variation of $\Omega_{0}(x,p)$ and hence
influence the global mode structure. 
The simulations presented here are based upon a global extension to the CYCLONE base
case and employ the circular Miller equilibrium model. 
In an equilibrium with radially varying profiles of $a/L_{T}$ and $a/L_{n}$, peaked at $x=0$, 
and with all other equilibrium profiles held constant, including $\eta_{i} = L_n/L_T$, 
$\Omega_{0}(x,p)$ is found to have a stationary point.
The reconstructed global mode sits at the outboard
mid-plane of the tokamak, with global growth rate,
$\gamma\sim$Max$\left[\gamma_{0}\right]$. Including the radial variation of 
other equilibrium profiles like safety factor and magnetic shear, 
leads to a mode that peaks away from the outboard mid-plane, with a reduced global growth rate. 
Finally, the influence of toroidal flow shear has also been investigated through the introduction of a Doppler
shift, $\omega_{0} \rightarrow \omega_{0} - n\Omprime x$, where 
$\Omega_{\phi}$ is the equilibrium toroidal flow, and a prime denotes the radial derivative. 
The equilibrium profile variations introduce an asymmetry into the global growth rate spectrum with respect to the
sign of $\Omprime$, such that the maximum growth rate is achieved with
non-zero shearing, consistent with recent global gyrokinetic calculations.
\end{abstract}

~~~~~~~
~~~~~~~
\section{Introduction}
\label{introduction}

Tokamaks \cite{Wesson_1} provide one of the most stable and promising
configurations for magnetic confinement fusion. However, confinement in tokamaks
is not perfect; there are a number of mechanisms by which energy and particles
can be transported across the magnetic flux surfaces from the core confinement
region to the plasma edge. The main contribution is typically due to turbulent
transport, which is widely believed to originate primarily from
microinstabilities driven by density and temperature gradients. These ``drift''
modes, with low frequency compared to the cyclotron frequency, are the dominant
tokamak microinstabilities \cite{Wesson_1, Horton_1}, and the turbulence they
drive influences the minimum size of magnetic confinement fusion reactors, such
as ITER and DEMO \cite{ITER}. It is therefore important to understand these
drift modes in order to seek ways to reduce their impact. Previous theoretical
and numerical studies have shown that flow shear can control the stability of
drift waves, providing a mechanism to suppress or even stabilise them completely 
\cite{Connor_0, Terry_1, Kishimoto_1, Roach_1, Biglari_1, Waltz_1}. 
While microinstabilities must ultimately be saturated by nonlinear physics (e.g. zonal flows), 
it nevertheless remains important to understand the structure of linear instabilities and the threshold gradients 
associated with their onset: linear physics lies close to the heart of  
many simplified plasma models that are of considerable value, e.g. the quasi-linear based TGLF model of 
anomalous transport in tokamaks \cite{Staebler_POP2007}.
Microinstabilities are often investigated via
numerical solutions of the gyrokinetic equation \cite{Rutherford_1, Frieman_1, Taylor_0}
and several different approaches are typically used, as we now
discuss. 

For high toroidal mode numbers, $n$, the rational flux surfaces (i.e. those
where the safety factor, $q$, is rational) are closely packed so equilibrium
quantities vary only weakly from one surface to the next. Two length scales can
then be identified: the equilibrium length scale, characterised by the minor
radius, $a$, and the distance between rational surfaces, $\Delta= (n \; dq/dr)^{-1}$,
where the derivative is with respect to radius, $r$. The separation
between these length scales is exploited in ballooning theory \cite{Taylor_1,
Dewar_0, Rewoldt_1, Connor_2} to perform an expansion in the small parameter, $\Delta/a$. The
rational surfaces spanned by a mode are then equivalent to leading order. Local
gyrokinetic codes, like GS2 \cite{Rewoldt_1, Dorland_1}, exploit this
``ballooning symmetry'' to reduce the intrinsic 2D problem (in radius, $r$, and
poloidal angle, $\theta$) to 1D in the extended ballooning coordinate, $\eta$,
which is aligned with magnetic field lines. Local gyrokinetic codes only
provide the structure along the magnetic field line, together with the local
eigenvalue, $\Omega_0(x,p)$ (where $x=\left(r-r_{0}\right)/a$ is the 
normalised radial distance from the reference surface, $r_0$, and the ballooning phase angle,
$p$, is the value of $\eta$ where the radial derivative of the eigenfunction is zero). 
Local codes, however, provide neither the radial mode structure nor the global eigenvalue $\Omega$, as
both $x$ and $p$ are free parameters at this lowest order in the ballooning expansion where local gyrokinetics is valid. 
The radial mode structure and global eigenvalue, $\Omega$, are determined at the next order in the $\Delta/a$ expansion, where the eigenfunction dependence on $x$ and $p$ 
becomes constrained. 

Exploiting the higher order theory to solve for the full global eignmode has been demonstrated previously for a simplified fluid model of ion temperature gradient (ITG) modes \cite{Dickinson_1}. In this paper, we build on \Ref{Dickinson_1} to show how this formalism can be extended to more realistic full gyrokinetic plasma models. Our approach exploits the GS2 code \cite{Rewoldt_1,
Dorland_1} to provide the local mode structure, $\xi(x, p, \eta)$, and $\Omega_{0}(x,p)$, from which the higher order theory provides the global mode structure and $\Omega$. 

It has previously been demonstrated that under the very special conditions where $\Omega_{0}(x,p)$ has
a stationary point, a so-called isolated or conventional ballooning mode is
obtained from the higher order analysis \cite{Taylor_1}. This type of mode,
originally studied in the context of ideal MHD theory \cite{Taylor_2, Connor_1}, is
usually located at the outboard mid-plane, where the poloidal angle $\theta=0$ \cite{Dickinson_1}. The higher order theory for such an isolated mode provides a global complex mode frequency $\Omega$, which includes an $O(1/n)$ correction to the local eigenvalue, $\Omega_{0}(x,p)$, evaluated at the maximally unstable $x$ and $p$. This maximally unstable ''isolated mode'' requires a stationary point in the local complex mode frequency $\Omega_{0}(x,p)$, which is a very special situation. More usually a second class of mode, known as the ``general mode'', would be expected in most situations. The higher order theory for these general modes predicts that they peak away from the outboard mid-plane and have a reduced growth rate relative to the isolated mode. Local gyrokinetic simulations alone cannot distingush between these two types of mode, but evidence for both can be found in global gyrokinetic simulations that include the radial variation of equilibrium profiles. For example, electrostatic ITG modes are found to be 
shifted relative to the outboard mid-plane in linear global gyrokinetic simulations of ASDEX Upgrade plasmas \cite{Bottino_1}. 
Such up-down asymmetries are important as they could provide a mechanism for flow generation in tokamaks \cite{Camenen_1, Peeters_0, Diamond_2013, Parra_2014}, which may be important in a machine like ITER for which the external torque is small. While nonlinear simulations are likely neccessary for a complete understanding of turbulence and flows in plasmas, linear theory provides a picture of the important physical mechanisms. The technique we present here is an alternative approach to full global simulations that uses an efficient formalism to shed light on the key linear physics.

The paper is organised as follows. In Section \ref{technique} we review the
theoretical formalism on which this paper is based. Results from applying
this technique to the so-called CYCLONE base case \cite{Dimits_1, Falchetto_0}, are presented in Section \ref{results}.
Employing this standard test case enables us to compare with previously published global simulations \cite{P.Hill_1}. Finally, Section \ref{conclusion} summarises our conclusions, and plans for
future work.

~~~~~~~
~~~~~~~
~~~~~~~
\section{The technique: From local to global gyrokinetic
calculations}\label{technique}
We employ the initial value local gyrokinetic code, GS2 \cite{Rewoldt_1,
Dorland_1}, to solve the linearised gyrokinetic equation \cite{Rutherford_1,
Frieman_1} numerically. GS2 provides the local eigenvalue, $\Omega_{0}(x,p)$ as well as the mode structure along a given magnetic field line $\xi(x,p,\eta)$ in the infinite domain in $\eta$ with a periodic dependence on $p$. 
To reconstruct the linear global mode properties from these local modes, we employ the Fourier-Ballooning (FB) representation \cite{Mahajan_1}:
\begin{equation}
\label{equ_1}
\phi(x,\theta,t) = \int_{-\infty}^{\infty}\xi(x, p, \theta,
t)\exp(-inq_{0}\theta)\exp[-in\qprime x(\theta - p)]A(p) dp
\end{equation}
where $\phi(x,\theta,t)$ is the global mode structure for fluctuations in the electrostatic potential, $x$ is the radial variable, $q_{0}$ is the safety factor at $x=0$, and the local mode structure $\xi(x,p,\theta,t)$ is obtained from GS2. Note that, the $x$-dependence in $\xi$ is due to a slow dependence of the equilibrium on $x$, and is a parameter at this order. For the linear global modes studied in this paper, $\phi(x,\theta,t)$ and $\xi(x,p,\theta,t)$ have a separable
time dependence of the form $\propto e^{-i \Omega t}$ and $e^{-i \Omega_{0}(x,p) t}$ where $\Omega$ and $\Omega_{0}(x,p)$ are the global and local complex linear eigenmode frequencies, respectively. In the rest of the paper,
we will frequently drop the explicit $t$ dependence, and use $\phi(x, \theta)$ and $\xi(x, p, \theta)$ to denote the separated spatial dependent 
part of the linear eigenmode structure.
The mapping in \Eqn{equ_1} from the infinite domain in $\eta$ to the periodic poloidal angle $\theta$, is possible because of 
the symmetry property $\xi(x, p+2l\pi,\theta+2l\pi)=\xi(x,p,\theta)$, for any integer $l$.

In order to employ \Eqn{equ_1} to find the global mode structure, $\phi(x,\theta,t)$, we must evaluate the envelope
$A(p)$, which can be obtained from the higher order theory as follows. We seek a global eigenmode satisfying
\begin{equation}
\label{equ_1a}
\frac{\partial\phi(x,\theta, t)}{\partial t} = -i \Omega \phi(x,\theta, t)
\end{equation}
where $\Omega=\omega+i\gamma$,
with $\omega$ and $\gamma$ corresponding to the global frequency and growth rate,
respectively. Substituting \Eqn{equ_1} into \Eqn{equ_1a}, and noting
$\partial\xi(x,p,\theta,t)/\partial t = -i \Omega_0(x,p) \xi(x,p,\theta,t)$, gives
\begin{equation}
 \label{equ_1b}
\Omega \phi(x,\theta, t) = \int_{-\infty}^{\infty}
\Omega_0(x,p)\xi(x,p,\theta,t)\exp(-inq_{0}\theta)\exp[-in\qprime{}x(\theta -
p)]A(p) dp
\end{equation}
Using \Eqn{equ_1} for $\phi$, we can then write our eigenmode condition in the
form:
\begin{equation}
\label{equ_1c}
\int_{-\infty}^\infty \left[\Omega - \Omega_{0}(x,p)\right] \xi(x,p,\theta,t)
\exp(-inq_{0}\theta) \exp[-inq^\prime{}x(\theta - p)] A(p)dp=0
\end{equation}
Here, the local complex mode frequency, $\Omega_{0}(x, p)$ and $\xi(x,p,\theta,t)$ are mapped out by
running the local gyrokinetic code, GS2, several times spanning the required range of
$x$ and over $-\pi \leq p \leq \pi$. This process is trivially parallelised, and therefore will usually be more efficient than a full global simulation. 
Anticipating radially localised solutions in $x$ we may Taylor expand $\Omega_0$ to second order in $x$ about $x=0$ and Fourier expand in $p$
to write:
\begin{equation}
\label{equ_2}
\Omega_{0}(x,p)=\sum_{k=0}^{N}\sum_{m=0}^{2}a_{k}^{(m)}x^{m}cos(kp)
\end{equation}
where $N$ is the number of Fourier modes retained. The coefficients
$a_{k}^{(m)}$ are complex numbers, obtained by fitting the expansion to the full $\Omega_{0}(x,p)$
results which are obtained from GS2. Now substituting \Eqn{equ_2} into \Eqn{equ_1c}, we obtain
\begin{eqnarray*}
\int_{-\infty}^\infty & \left[\Omega - \sum_{k=0}^{N} \left( a_{k}^{(0)} + a_{k}^{(1)}x + a_{k}^{(2)}x^2 \right) \cos(kp) \right]A(p) \\ & \xi(x,p,\theta,t) \exp(-inq_{0}\theta)\exp[-in\qprime{}x(\theta - p) ] dp  = 0.
\end{eqnarray*}
We note that $n\qprime p$ is a radial wave number. Therefore, following standard Fourier transform procedures and assuming $A(p)$ varies much more rapidly than $\xi$ with $p$, allows us to replace $x^{m} A(p)$ with $(-i/n\qprime{})^{m} d^{m}A/dp^{m}$, to give:
\begin{eqnarray}
\label{equ_2a}
 \int_{-\infty}^\infty & \left[ \Omega A - \sum_{k=0}^{N}\left(a_{k}^{(0)} \cos(kp) -
\frac{ia_{k}^{(1)}}{n\qprime{}} \cos(kp) \frac{d}{dp}-\frac{a_{k}^{(2)}}{(n\qprime{})^2}
 \cos(kp) \frac{d^{2}}{dp^{2}}\right)A\right] \nonumber \\
& \xi(x,p,\theta,t) \exp(-inq_{0}\theta)\exp[-in\qprime{}x(\theta - p) ] dp = 0.
\end{eqnarray}
Equation (\ref{equ_2a}) must hold for all $x$ and $\theta$, which then provides our final equation for $A(p)$:
\begin{equation}
\label{equ_4}
\sum_{k=0}^{N_k} \left[-\frac{a_{k}^{(2)}}{(n\qprime{})^2}\cos(kp)\frac{d^{2}}{dp^{2}} - \frac{ia_{k}^{(1)}}{n\qprime{}}\cos(kp)\frac{d}{dp}+a_{k}^{(0)}\cos(kp)\right]A= \Omega A
\end{equation}
As $\phi(x,\theta,t)$ must be periodic in $\theta$, the representation \Eqn{equ_1} requires $A(p)$ to be periodic in $p$. Thus \Eqn{equ_4} must be solved with periodic boundary conditions to determine the envelope $A(p)$ and the global
complex mode frequency, $\Omega$ as an eigenvalue. Knowledge of $A(p)$, together
with $\xi(x, p , \theta)$ from GS2 then allows the full 2D eigenfunction, $\phi(x,
\theta)$, to be reconstructed from \Eqn{equ_1}. 

Equation (\ref{equ_4}) can be solved analytically in the limiting cases where either $a_{k}^{(1)}=0$,
or where $a_{k}^{(2)}=0$ for all $k$; the former implies that $\Omega_{0}(x, p)$ is
stationary at $x=0$, i.e. $\partial \Omega_0 /\partial x |_{x=0} = 0$. These two
limits lead to two different classes of eigenmode \cite{Taylor_1, Dickinson_1}
which are referred to as ``isolated modes'' in the special case where
$a_{k}^{(1)} = 0$, and ``general modes'' in the more usual situation where
$a_{k}^{(1)} \neq 0$ (when the terms in $a_k^{(2)}$ can be neglected in the
limit of large $n$). In this paper we shall retain both terms and solve
\Eqn{equ_4} numerically.

\section{Results}	\label{results}

\begin{figure}[t!]
		\centering
\includegraphics[width=1.0\textwidth]{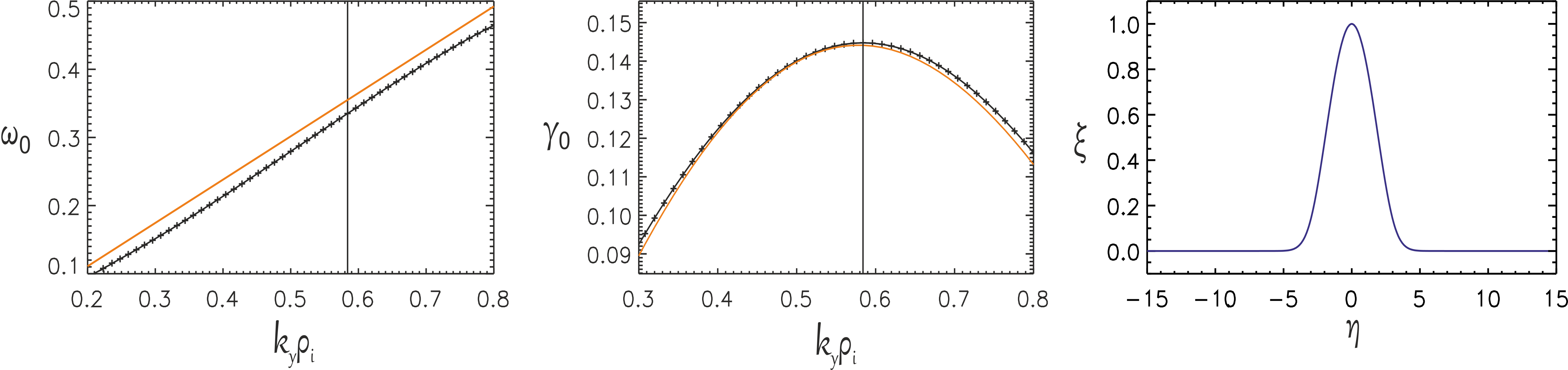}
\caption{\label{fig_1}From left to right: The variation of (a) real frequency,
$\omega_{0}$, (b) linear growth rate, $\gamma_{0}$, with $k_{y} \rho_{i}$, for
the dominant modes at $p=0$ for two values of the normalised 
ion collision frequency, $\nu_{ii}a/c_s=0$ (solid line) and $\nu_{ii}a/c_s=0.28$ (solid line with $*$ symbols); (c) shows the local mode structure, $\xi(x=0, \eta, p=0)$, for the
most unstable mode at $k_{y} \rho_{i}=0.58$, as a function of ballooning coordinate $\eta$ along the magnetic field line. 
Note that both $\omega_{0}$ and $\gamma_{0}$ are measured in units of
$(c_{s}/a)$ and these local simulations have been carried out at mid-radius, i.e. $r=a/2$.}
\end{figure}

 \begin{SCfigure}
		\centering
 \includegraphics[width=0.50\textwidth]{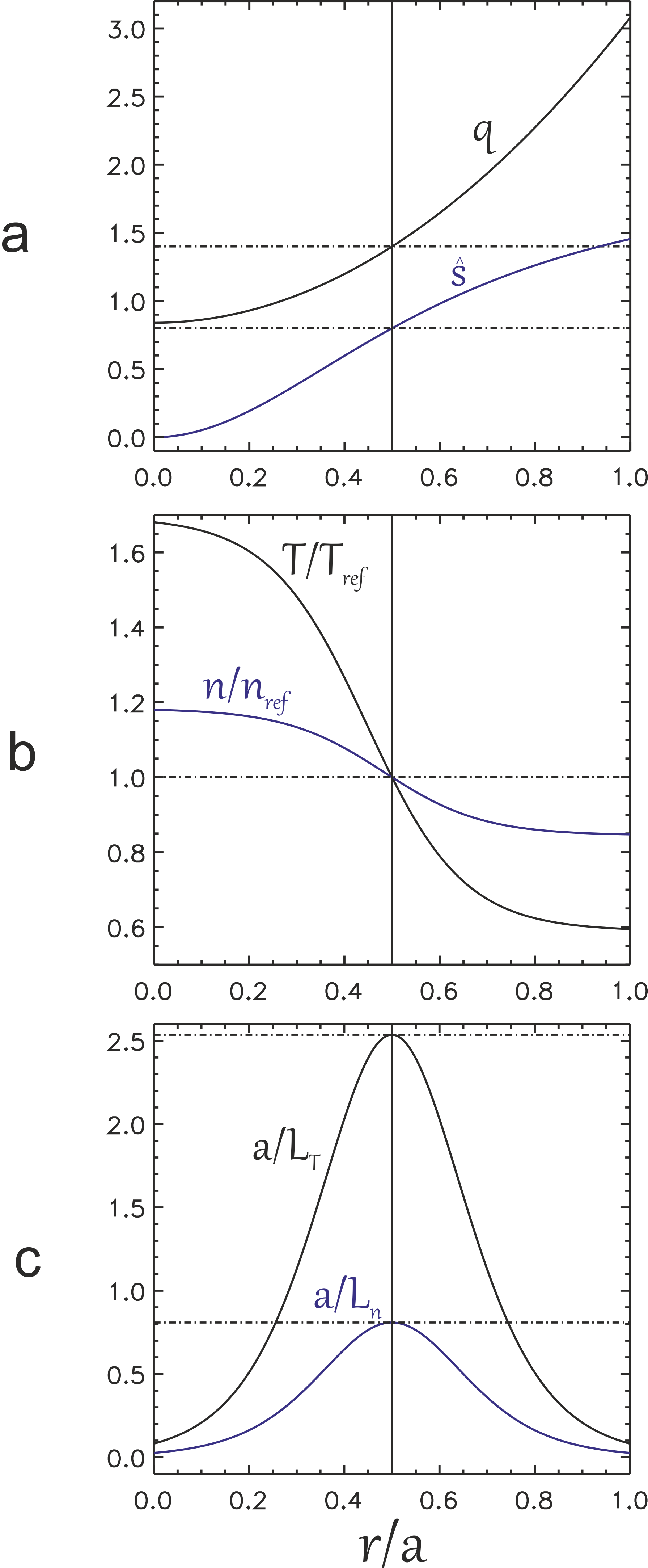}
 \caption{The radial profiles used in this
paper, chosen to match \Ref{P.Hill_1}. (a) Safety factor, $q=0.84+2.24(r/a)^{2}$ and magnetic shear
$\shat = 2\left(1-\frac{0.84}{q}\right)$ and (b) temperature, $\frac{T}{T_{ref}}=\exp \left[-\Delta
T\left(\frac{a}{L_{T}}\right)tanh(\frac{x}{\Delta T})\right]$, and density
$\frac{n}{n_{ref}}=\exp \left[-\Delta n\left(\frac{a}{L_{n}}\right)tanh(\frac{x}{\Delta n})\right]$ and finally (c) The
temperature, $\frac{a}{L_{T}}=\left(\frac{a}{L_{T}}\right)_{0}\left(cosh(\frac{x}{\Delta T})\right)^{-2}$, and density, 
$\frac{a}{L_{n}}=\left(\frac{a}{L_{n}}\right)_{0}\left(cosh(\frac{x}{\Delta n})\right)^{-2}$, gradients. 
Here, $\Delta T = \Delta n = 0.208$. The temperature, $\left(\frac{a}{L_{T}}\right)_{0}=2.54$ and density
$\left(\frac{a}{L_{n}}\right)_{0} = 0.8$ gradients are evaluated at the centre
of the domain ($r=r_{0}=0.5a$). Note that, the horizontal dashed lines
represent the value of a particular quantity at $x=0$ ($r=r_{0}$), where $x=\frac{r-r_{0}}{a}$.}
\label{fig_2}
\end{SCfigure}
	
Before considering the global calculations, we first describe the local
ballooning analysis. Our focus is on ITG modes, believed to be one of the
common causes for turbulent transport in many tokamak plasmas. Nevertheless, we note that our technique applies to all classes of high $n$ toroidal drift modes. 
We adopt the CYCLONE parameters \cite{Dimits_1, Falchetto_0} and for convenience we have limited
ourselves to electrostatic calculations with adiabatic electrons, which allows us to make comparisons with previously published results from global simulations. The model parameters are given in Table \ref{table_1} for a Miller equilibrium with circular flux surfaces. 

\Fig{fig_1} shows the local real frequency, $\omega_{0}$, and growth rate, $\gamma_{0}$, obtained from GS2 as functions
of normalised binormal wavenumber, $k_{y}\rho_{i}$, for the dominant modes with $p=0$, together with the local mode structure, $\xi$ at $x=0$. The results for a normalised ion-ion collision frequency of $\nu_{ii}a/c_s=0.28$ are found to be very similar to those for a collisionless plasma for $p=0$. We use this finite collision rate in the remaining calculations as it helps damp unphysical modes found by GS2 at values of $p$ close to marginal stability, and yet gives local eigenmode result similar to the collisionless case. The most unstable mode is found at $k_{y}\rho_{i}=0.58$ which corresponds to toroidal mode number $n=39$. Now we apply the technique presented in Section \ref{technique} to reconstruct the global structure for the most unstable mode.
\begin{table}[t] 
\caption{CYCLONE equilibrium parameters.}
\centering 
\begin{tabular}{|c|c||c|c|}\hline
Parameter &	Value & Parameter & Value \\\hline	
  \shat		        & 	\multicolumn{1}{c||}{0.8} 	 &  $r_{0}(m)$	      									& \multicolumn{1}{c|}{0.313}\\\hline
  $q_{0}$		      & 	\multicolumn{1}{c||}{1.4} 	 &  $\beta$           									& \multicolumn{1}{c|}{0.0}\\\hline
  $a/L_{T}$	      & 	\multicolumn{1}{c||}{2.54}   &  $n\qprime$        									& \multicolumn{1}{c|}{144}\\\hline
  $a/L_{n}$	      & 	\multicolumn{1}{c||}{0.81}   &  $\nu_{ii} a/c_s$										& \multicolumn{1}{c|}{0.28}\\\hline
  $k_{y}\rho_{i}$ & 	\multicolumn{1}{c||}{0.58}   &  $\frac{T_{i}}{T_{e}}$								& \multicolumn{1}{c|}{1.0}\\\hline 
	$a(m)$	        & 	\multicolumn{1}{c||}{0.625}  &  $\rho_{i} (m)$ 											& \multicolumn{1}{c|}{0.003384}\\\hline
  $R(m)$    	    & 	\multicolumn{1}{c||}{1.70} 	 &  $\rho_{\star} =\frac{\rho_{i}}{a}$	& \multicolumn{1}{c|}{0.005415}\\\hline
\end{tabular} 
\label{table_1} 
\end{table}

\begin{table}[t] 
\caption{The model coefficients, $a_{k}^{m}$, with ten Fourier modes. The real and
imaginary components contribute to the real frequency, $\omega_{0}$, and linear
growth rate, $\gamma_{0}$, respectively. Note that coefficients with $m=1$ are
all zero for this special case where only profiles in $a/L_{T}$ and $a/L_{n}$ are retained.}
\centering 
\begin{tabular}{|c|c|c|}\hline
\backslashbox{k}{m} &	0 & 2 \\\hline	
  0		&	  \multicolumn{1}{c|}{ 0.1177 - 0.0680 i} & \multicolumn{1}{c|}{-1.5689 - 1.9352 i}\\\hline
  1		& 	\multicolumn{1}{c|}{ 0.1804 + 0.1221 i} & \multicolumn{1}{c|}{ 1.3347 - 2.2466 i}\\\hline
  2		& 	\multicolumn{1}{c|}{ 0.0462 + 0.0461 i} & \multicolumn{1}{c|}{ 0.0825 - 0.8734 i}\\\hline
  3		& 	\multicolumn{1}{c|}{ 0.0229 + 0.0231 i}	& \multicolumn{1}{c|}{-0.0015 - 0.2477 i}\\\hline
  4		& 	\multicolumn{1}{c|}{ 0.0068 + 0.0098 i} & \multicolumn{1}{c|}{-0.1007 - 0.0227 i}\\\hline
  5		& 	\multicolumn{1}{c|}{ 0.0012 + 0.0078 i} & \multicolumn{1}{c|}{-0.1090 + 0.1078 i}\\\hline
	6		& 	\multicolumn{1}{c|}{-0.0022 + 0.0045 i} & \multicolumn{1}{c|}{-0.1134 + 0.1240 i}\\\hline
	7		& 	\multicolumn{1}{c|}{-0.0033 + 0.0023 i} & \multicolumn{1}{c|}{-0.0861 + 0.0856 i}\\\hline
	8		& 	\multicolumn{1}{c|}{-0.0035 - 0.0000 i} & \multicolumn{1}{c|}{-0.0461 + 0.0105 i}\\\hline
	9		& 	\multicolumn{1}{c|}{-0.0026 - 0.0018 i} & \multicolumn{1}{c|}{ 0.0111 - 0.0525 i}\\\hline	
\end{tabular} 
\label{table_2} 
\end{table}

\subsection{Global Calculations: Isolated modes with flat
\texorpdfstring{$\eta_{i}$}{Lg} profile} \label{global1}

\begin{figure}[t!]
	\centering
	\includegraphics[width=1.0\textwidth]{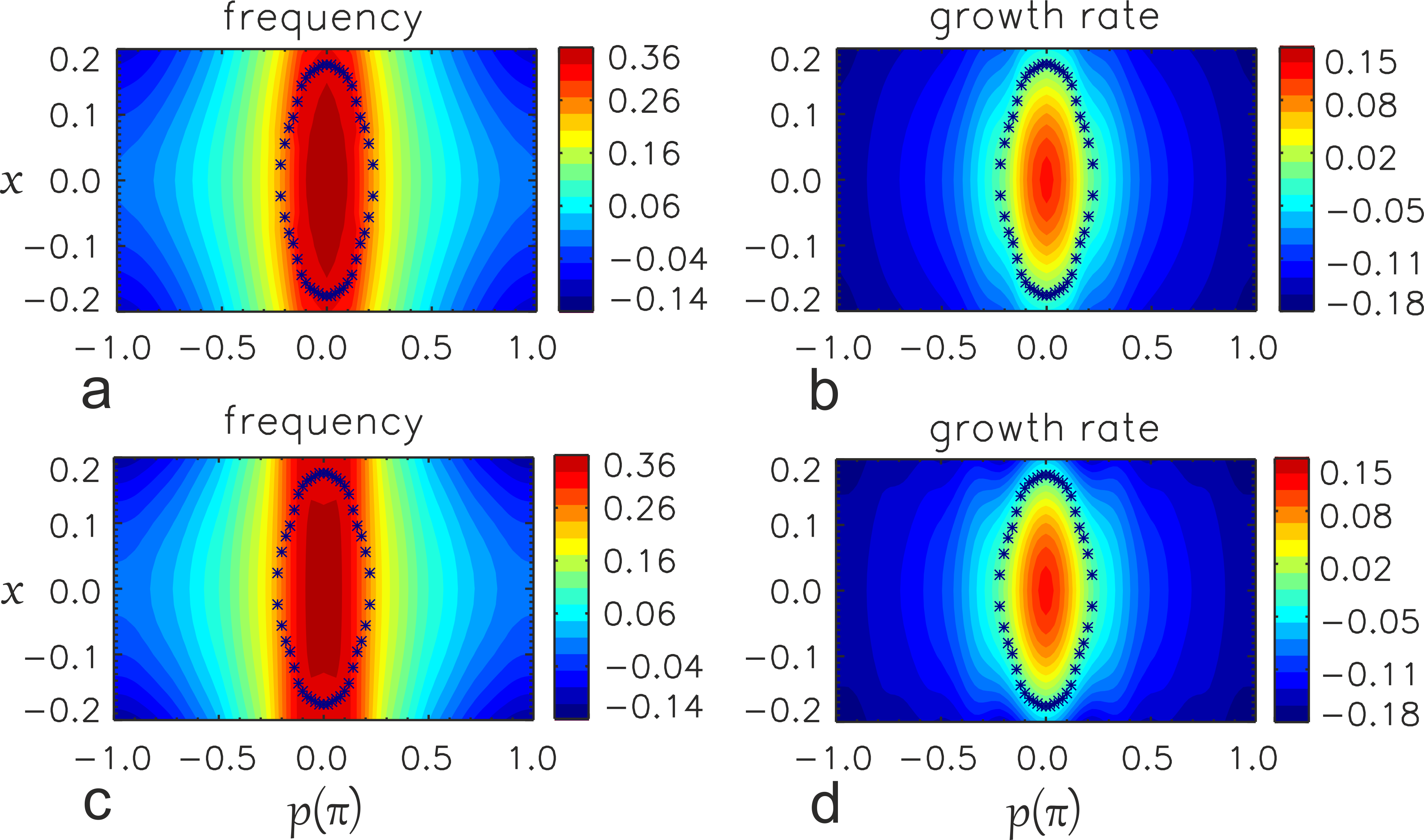}
\caption{\label{fig_3}Contour plots of real and imaginary parts of the local
complex mode frequency, measured in unit of $(c_{s}/a)$, as functions of
radius, $x$, and ballooning phase angle, $p$, for parabolic $L_{T}$ and $L_{n}$ radial
profiles (see \Fig{fig_2}c), while excluding other profile variations. (a)
and (b) are, respectively, the frequency and growth rate obtained from the local
gyrokinetic code, GS2. The corresponding frequency and growth rate from the
fitted model, using \Eqn{equ_2}, are presented in (c) and (d) respectively. The $\star$ symbols indicate
the marginal stability contour where $\gamma_{0}(x,p)=0$.}
\end{figure}	

\begin{figure}[t!]
	\centering
	\includegraphics[width=1.0\textwidth]{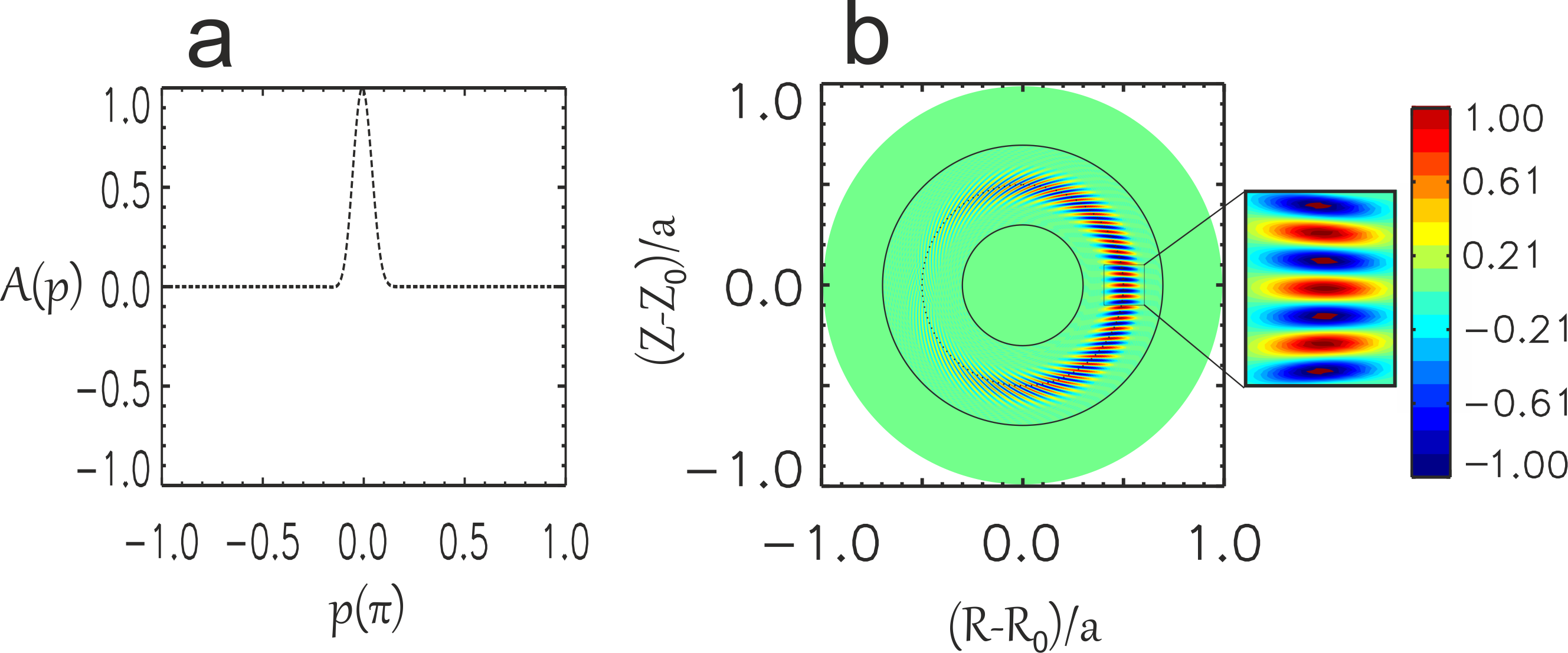}
\caption{\label{fig_4}(a) presents the numerical solution of
the envelope function, $A(p)$, obtained from \Eqn{equ_4}, as a
function of ballooning phase angle, $p$, and (b) shows the reconstructed electrostatic potential
global mode structure, $\phi(x, \theta)$, in the poloidal plane. The profile
variations other than $L_{T}$ and $L_{n}$ are excluded from this calculation, and the mode structure,
centred on the outboard mid-plane, is aligned radially where $\theta = 0$. The solid black lines indicate the radial domain of the calculation.}
\end{figure}

\begin{SCfigure}
	\centering
	\includegraphics[width=0.50\textwidth]{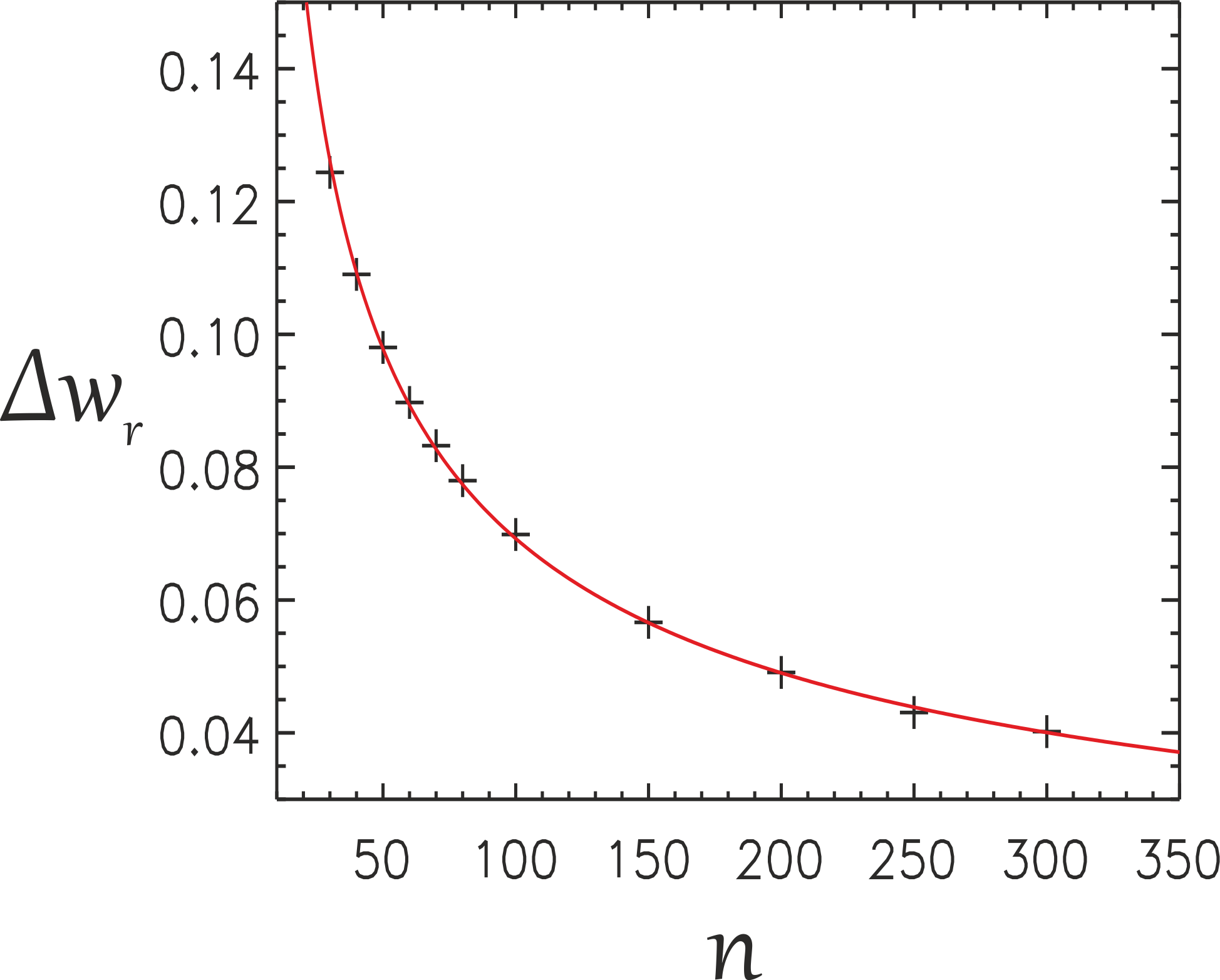}
	\caption{\label{fig_5} The radial mode width, $\Delta w_{r}$, as a
function of toroidal mode number, $n$. It scales, approximately, inversely with square root of $n$
according to: $\Delta w_{r} = 0.687 n^{-0.49}$, as expected for an isolated mode.}
\end{SCfigure}

	\begin{figure}[t!]
	\centering
	\includegraphics[width=1.0\textwidth]{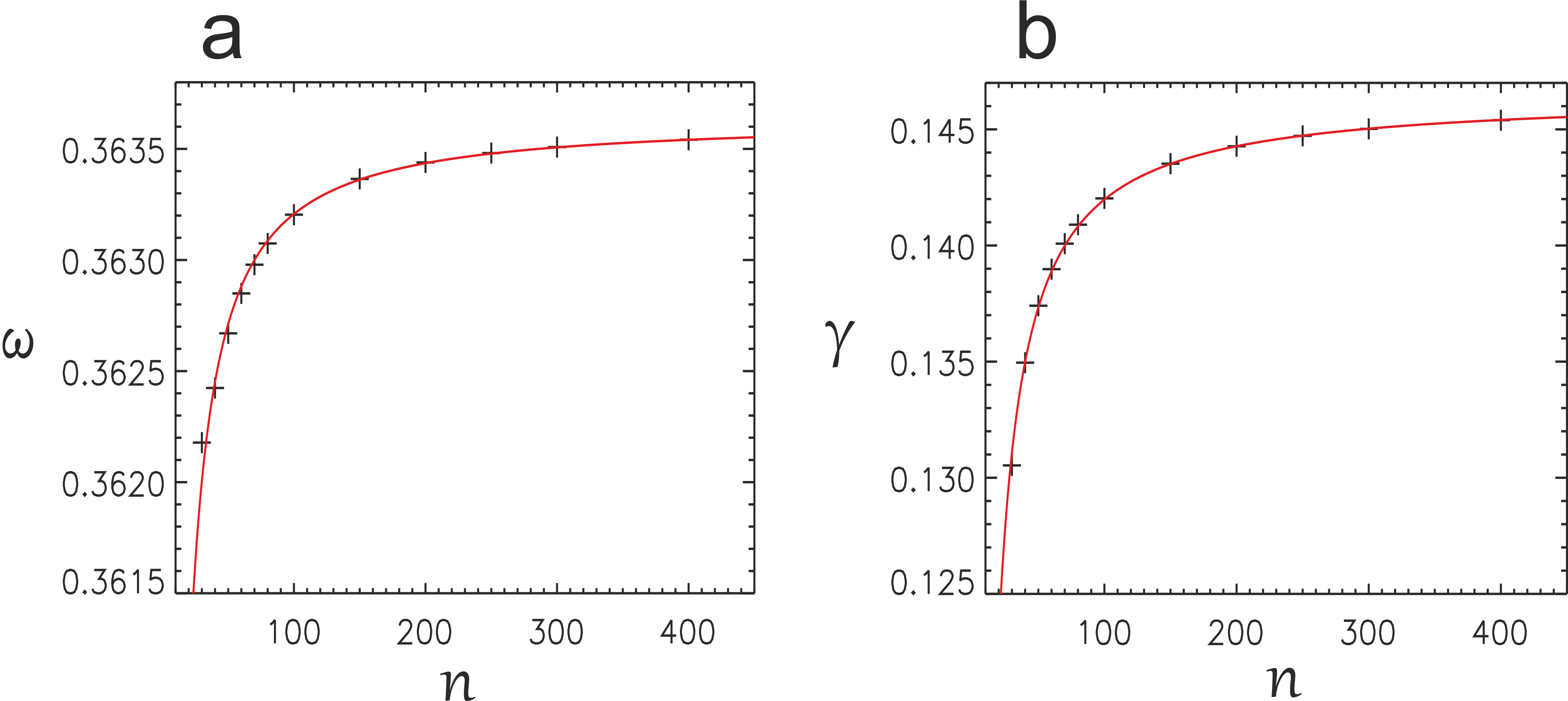}
\caption{\label{fig_6} (a) The global real frequency $\omega$ and (b) growth rate
$\gamma$ as functions of toroidal mode number, $n$, for fixed
$k_{y}\rho_{i}=0.58$. The $\omega$ and $\gamma$ are measured in units of
$(c_{s}/a)$. They scale with $n$ as: $\omega = 0.364 - 0.072 n^{-1.12}$,
$\gamma = 0.147 - 0.492 n^{-1.02}$.}  
\end{figure}

We first consider radially varying profiles of temperature and density, such that the gradients
$a/L_{T}$ and $a/L_{n}$ peak in the centre of the domain at $x=0$ while maintaining a flat profile in 
$\eta_{i}=L_{n}/L_{T}$ (see \Fig{fig_2}c).
Initially we fix all other equilibrium profiles to the constant values given in Table \ref{table_1} (ie. we use the dashed line profiles of \Fig{fig_2}a and b) 
\footnote{Results including the equilibrium variation shown in \Fig{fig_2}a and b will be considered in the next subsection.}.
GS2 calculations across the radial extent of the mode in $x$ and full range in ballooning phase angle $-\pi \leq p \leq \pi$ then provide the local complex mode frequency, $\Omega_{0}(x, p)$.
\Fig{fig_3} shows $\Omega_{0}(x,p)$ from GS2, compared to the fit using the
parameterisation given in \Eqn{equ_2}. The coefficients resulting from the fit are given in
Table \ref{table_2}.
As expected, because the linear drive profiles peak and are symmetric about $x=0$, the coefficients with $m=1$ are negligible and $\Omega_0$ has a stationary point at $x=0$, $p=0$. 

Solving \Eqn{equ_4} numerically with the values of $a_{k}^{m}$ in Table \ref{table_2} we obtain $A(p)$ and the associated eigenvalue $\Omega=0.362+0.135i$. Using the numerical solution for $A(p)$, and
$\xi(x,p,\theta)$ obtained from GS2, \Eqn{equ_1} provides the
global mode structure, $\phi(x, \theta)$. \Fig{fig_4} shows our solution for $A(p)$ and the
corresponding solution for $\phi(x,\theta)$ in the poloidal cross-section. We
see that $A(p)$ is localised, as is required for the procedure to be accurate (recall, we assumed $A(p)$ varies rapidly with $p$ to derive \Eqn{equ_2a}). As shown in \Fig{fig_4}, this leads to
a mode that balloons on the outboard mid-plane at $\theta=0$.

The radial mode width, $\Delta w_{r}$, and its variation with the toroidal mode
number, $n$ (or equivalently $1/\rho_{\star}$), has also been calculated and is
shown in \Fig{fig_5}. Here the width is defined as the full width half maximum
of a Gaussian fit to the magnitude of $\phi$ at $\theta=p_{0}$, where $p_{0}$
is the ballooning phase angle at which the global mode peaks poloidally. We find that
the radial width of the mode scales inversely with the square root of $n$,
$\Delta w_{r} \propto n^{-0.49}$ as expected for isolated modes \cite{Taylor_2}.

Changing the toroidal mode number (and $\rho_{\star}$ to keep $k_y \rho_i$ fixed) also affects the global mode frequency,
$\Omega$. \Fig{fig_6} shows that both the real frequency, $\omega$, and the
linear growth rate, $\gamma$, scale with $n$ according to: 
$\omega = 0.364 - 0.072 n^{-1.12}$ and $\gamma = 0.147 - 0.492 n^{-1.02}$, respectively. This
indicates that the finite $n$ correction scales inversely with the toroidal mode
number as expected from conventional ballooning theory \cite{Taylor_2}.  In the limit $n \rightarrow \infty$,
the global complex mode frequency $\Omega$ converges to the local complex
mode frequency at $x=p=0$, $\Omega_{0}(0,0)=\omega_{0}(0,0) + i\gamma_{0}(0,0)= 0.364 +
0.147 i$. This confirms that our results are consistent with analytic theory and higher 
order ballooning calculations presented in references \cite{Taylor_1, Dickinson_1, Romanelli_1}.

To sumarise, this mode has all the characterstics of the isolated mode identified in 
\cite{Taylor_1, Dewar_0, Dickinson_1}. It exists in this particular case because our choice of profiles ensure that $\Omega_0(x,p)$ has a stationary point at $x=0$, $p=0$, as is clear from \Fig{fig_3}. However, as we shall see in the following subsection, taking into account the radial variation of other equilibrium profiles introduces a
small, but significant, deviation from conventional ballooning (or isolated) modes.

~~~~~~~
\subsection{Global Calculations with Profile Variations}
\label{global2}
	
\begin{figure}[t!]
 \centering
\includegraphics[width=1.0\textwidth]{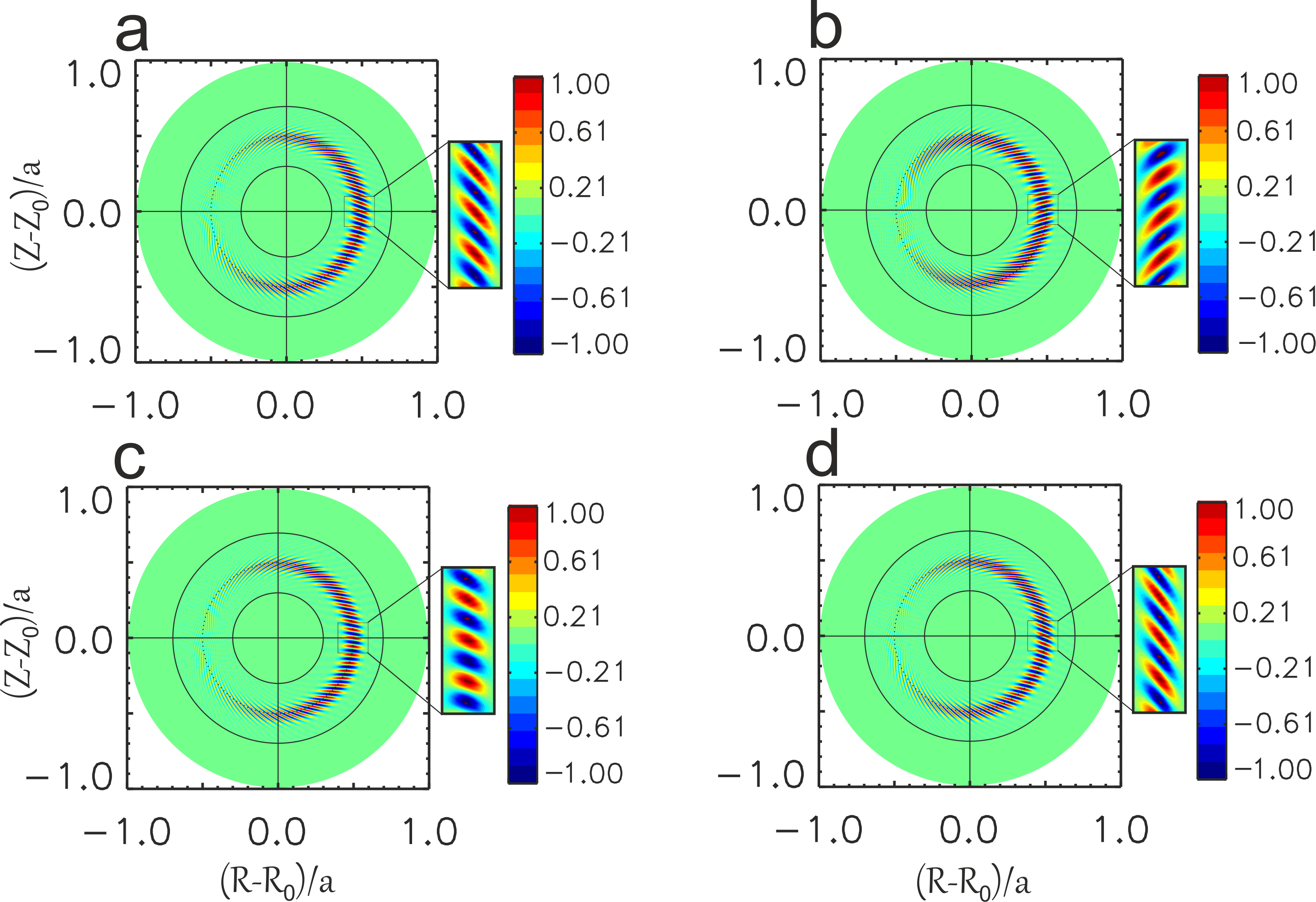}
\caption{\label{fig_7}The reconstructed electrostatic potential global mode
structure, $\phi(x, \theta)$, for $n=39$, in the poloidal plane for different
radial profile variations taken from \Fig{fig_2}: (a) $L_{T}$, $L_{n}$
and $x$ vary, here both temperature $T$ and density $n_e$ are assumed to be
constant, (b) $L_{T}$, $L_{n}$, $q$ and $\shat$ vary (c) $L_{T}$, $L_{n}$, $x$, $q$ and
$\shat$ vary and finally (d) full profile variation in which $L_{T}$, $T$,
$L_{n}$, $n_e$, $x$, $q$ and $\shat$ all vary.}
 \end{figure}

 
\begin{figure}[t!]
	\centering
	\includegraphics[width=0.5\textwidth]{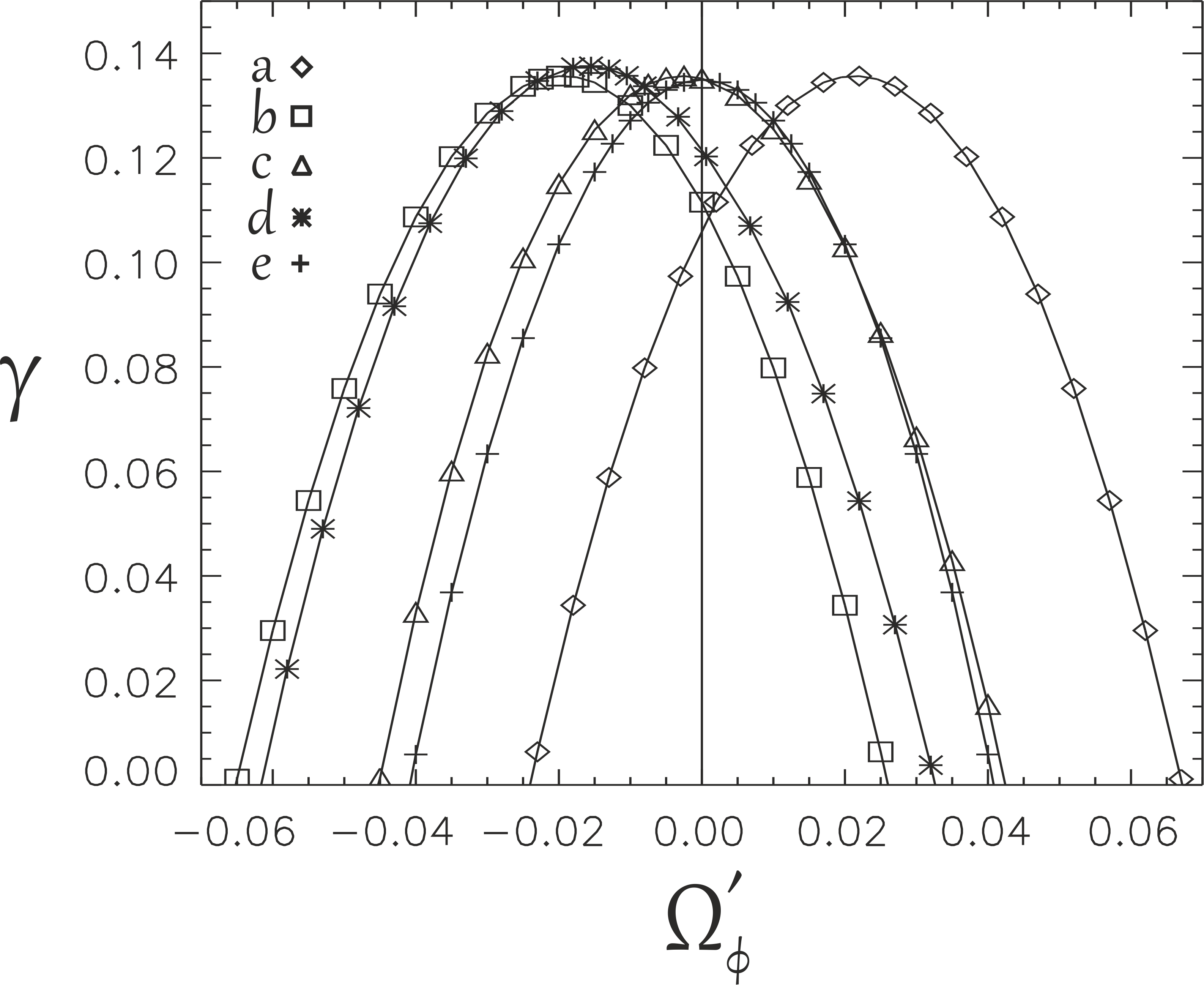}
	\caption{\label{fig_8}The linear growth rate, $\gamma$, measured in unit
of $(c_{s}/a)$, as a function of flow shear, $\Omprime$, calculated for the
most unstable mode with $k_{y}\rho_{i}=0.58$. The toroidal mode number, $n=39$
and $\qprime \approx 3.6$. For curve (e) the radial variation of the
profiles other than $L_{T}$ and $L_{n}$ are excluded, while the other curves
correspond to the profile variations of \Fig{fig_7}(a-d) respectively. }
\end{figure}

 
\begin{figure}[t!]
	\centering
	\includegraphics[width=1.0\textwidth]{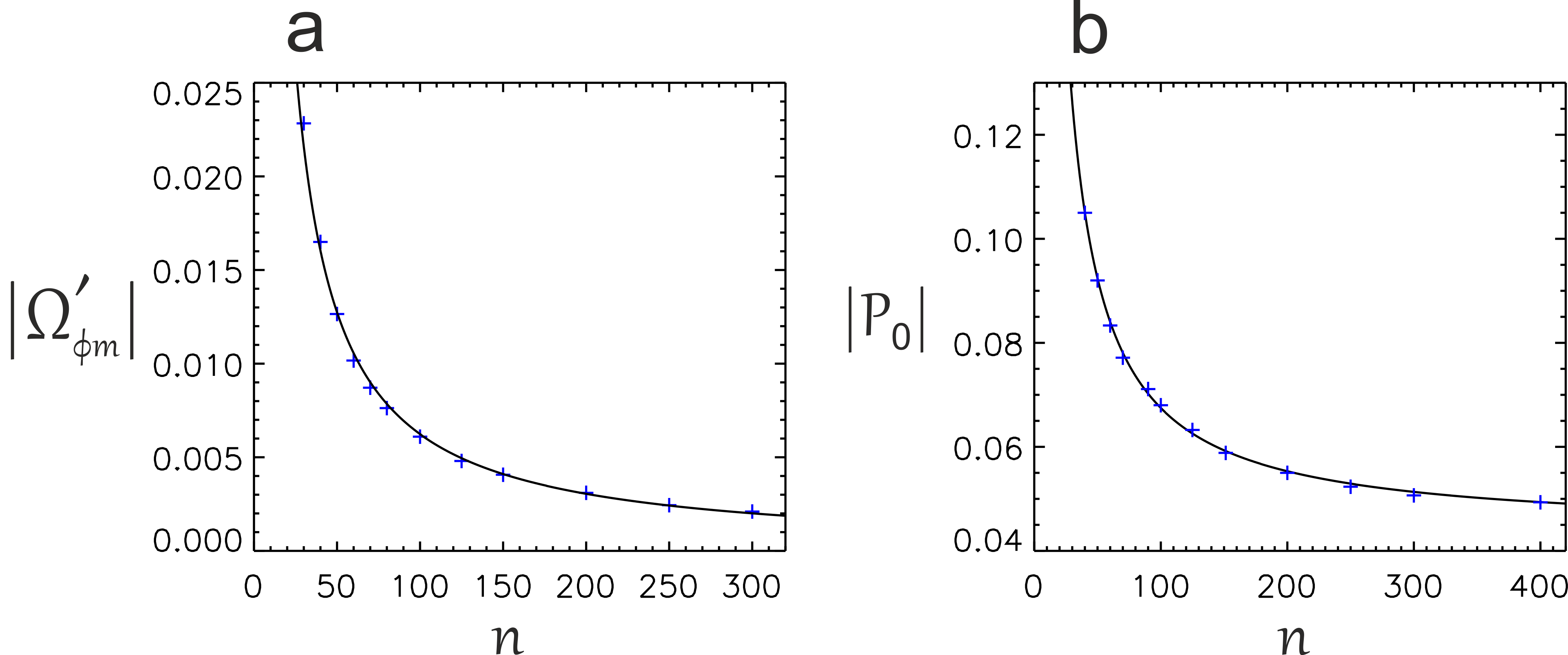}
	\caption{\label{fig_9}The scaling with toroidal mode numbeer, $n$, of (a) the value of flow shear
to get the most unstable mode (crosses), fitted to the curve $|\Omprimem| \sim 0.713n^{-1.03}$, and (b)
the offset in poloidal angle (crosses), fitted to the curve $\left|p_{0}\right| = 0.041 + 1.88n^{-0.925}$, 
for the full profile variation case, i.e curve$-$d
from \Fig{fig_8}.}
\end{figure}

\begin{figure}[t!]
	\centering
\includegraphics[width=1.0\textwidth]{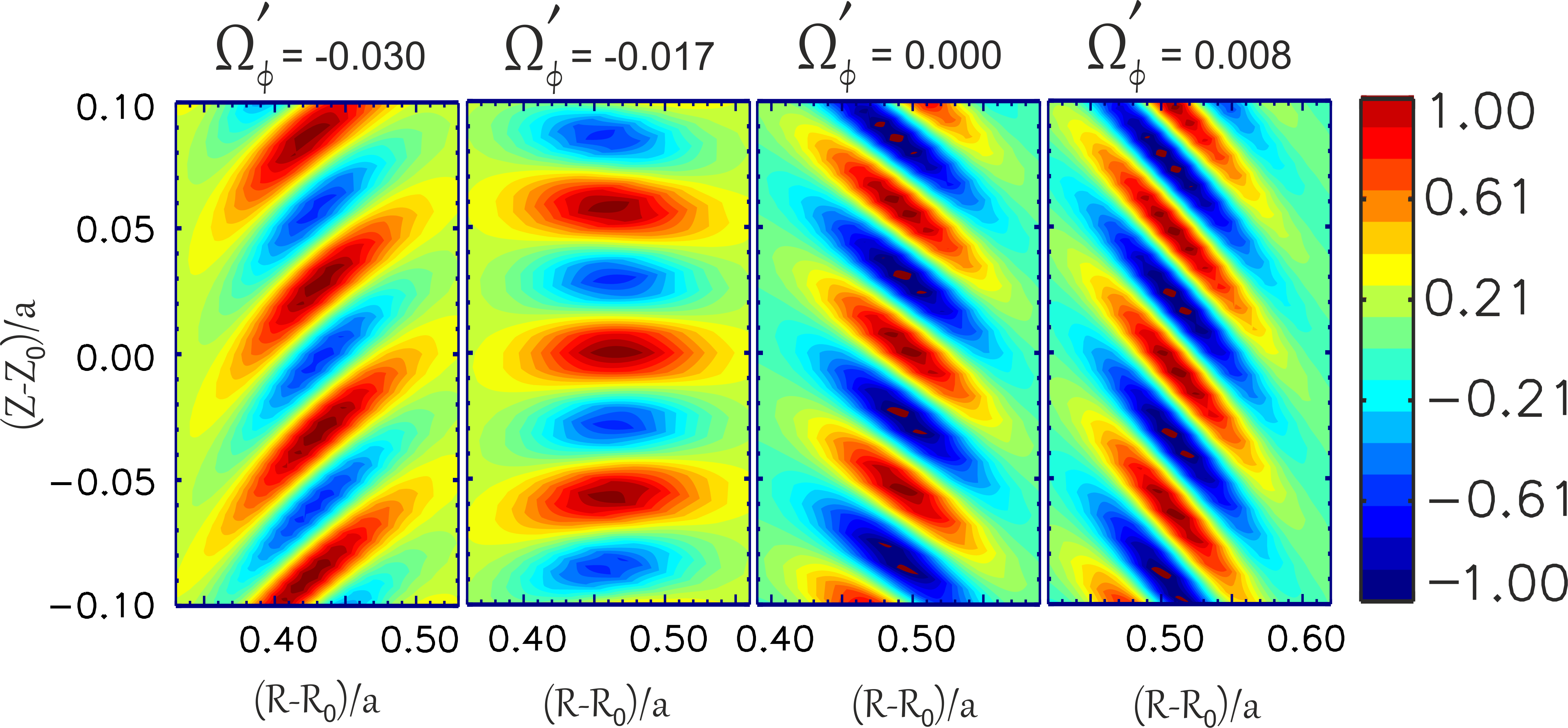}
	\caption{\label{fig_10}Reconstructed global mode structure, $\phi$, in a small
region of the poloidal cross-section at the outboard mid-plane (see \Fig{fig_4}), under the
combined effects of both radially varying equilibrium profiles from
\Fig{fig_7}-d and flow shear, $\Omprime$. From left to right,
$\Omprime=-0.03,-0.017, 0$ and $0.008$, respectively. For $\Omprime=0$,
the mode is already tilted, due to the profile variation effect, while a
critical value of flow shear, occurs at $\Omprime\approx -0.017$, which cancels out the
effect of profile variation and, once again, the mode structure is aligned
radially (as for a conventional ballooning mode). From \Fig{fig_8}, we see that
this is the maximally unstable flow shear.}
\end{figure}     

Now we introduce the other radially varying profiles given in \Fig{fig_2}, and
repeat the analysis of subsection \ref{global1}. These profiles vary over the radial scale of the
instability and give rise to a significant linear radial variation in
$\Omega_0(x,p)$, such that the $a_{k}^{(1)}$ terms in
\Eqn{equ_2} become significant. These extra linear terms impact $\omega_{0}(x,p)$ and
$\gamma_{0}(x,p)$ differently, which influences the global mode structure. Specifically, the positions of the maxima in
$\omega_{0}(x,p)$ and $\gamma_{0}(x,p)$ occur at different $x$, which affects the stability and poloidal position of the global mode. 

\Fig{fig_7} shows the reconstructed global mode structure for a number of cases
where we introduce different profiles. The effect of variation in radius $x$
itself is presented in panel (a). The mode shifts slightly downward with respect to the outboard mid-plane. Panel (b)
shows the effect of $q$ and $\shat$ profile variations and results in a mode
with an upwards poloidal shift. Panel (c) shows the combined influence of both (a) and (b) and demonstrates that for
this particular case they are competing and almost cancel. The result is slightly in favour
of the $x$ variation, leading to a slight net downward poloidal shift with respect to
the outboard mid-plane. Finally, in panel (d), in addition to the profiles
considered in (c), we include radial variations in the $T$ and $n_{e}$
profiles. The reconstructed global mode is then shifted poloidally
downward with respect to the outboard mid-plane. 

For the various profile variations considered above, the up-down poloidal
symmetry is broken. This moves the mode slightly off the mid-plane and towards the good curvature region, which reduces the growth rate compared to the flat
$\eta_{i}$ profile case considered in subsection~\ref{global1}. Specifically, for case $d$ with full profile variation we find $\Omega=0.334+0.119i$. As we can see, the profile variations tilt the structure on the outboard mid-plane. This finding agrees
qualitatively with other published results for full global simulations of linear
ITG modes \cite{P.Hill_1}. Our interpretation is that this is a consequence of profile variations in the equilibrium.

\subsection{Calculations with Profile Variations and Sheared Toroidal Flow}
We now consider the influence of a small toroidal flow shear on stability and the reconstructed
mode structures. This has been achieved by moving into the frame rotating 
toroidally with the plasma surface at $x=0$, and by introducing the following Doppler shift into 
$\Omega_{0}$ \footnote{While this method can in principle handle an arbitrary experimental profile of toroidal rotation, for 
the simple example here we assume that the toroidal flow varies linearly in $x$.}:
\begin{equation}
\label{equ_5}
\Omega_{0}(x,p) \rightarrow \Omega_{0}(x,p) -  n\Omprime x = \left[\omega_{0}(x,p)
- n\Omprime x\right] + i\gamma_{0}(x,p),
\end{equation} 
where $\Omprime$ is constant and sets the flow shearing rate. Thus $\Omprime
x$ represents the toroidal rotation frequency (normalised to $c_{s}/a$) of the magnetic flux surfaces measured
relative to the mode rational surface at $x=0$ where the reconstructed global
mode sits. We have considered $-0.06 \leq \Omprime \leq +0.06$. Note that we assume a purely toroidal flow here, 
as poloidal flows are expected to be damped in a tokamak plasma. From \Eqn{equ_5} it is clear that the flow shear influences the $a_{k}^{(1)}$
coefficients of \Eqn{equ_2}. Flow shear will therefore influence global modes in a similar way to the variations in other profiles, 
which we will now quantify.

Depending on the sign of the flow shear the reconstructed global mode shifts
upward or downward relative to the outboard mid-plane with a structure in the
poloidal plane that is very similar to that in \Fig{fig_7}. For a mode peaking
at $\theta=0$ in the absence of flow shear, increasing flow shear tilts the mode
structure on the outboard mid-plane and lowers the linear growth rate. \Fig{fig_8} shows how the growth rate varies with flow shear for
the different profile variations considered earlier in \Fig{fig_4} and
\Fig{fig_7}. Keeping only the symmetric $L_{T}$ and $L_{n}$ profiles the growth
rate curve is symmetric about $\Omprime=0$, shown by curve (e) of \Fig{fig_8}.
However, when other profile variations are taken into account, an asymmetry is
introduced into the dependence of growth rate on the sign of $\Omprime$, as
shown in \Fig{fig_8} (for curves a-d).
Only in the case where we have purely symmetric profiles about $x=0$ is the growth rate maximised at $\Omprime=0$. 
In the other cases the peak in growth rate occurs at non-zero flow shear. 
We can explain this as follows. We expect this maximally unstable isolated mode to exist when both $\omega_{0}(x,p)$ and $\gamma_{0}(x,p)$ are stationary
at the same position. From symmetry we anticipate $\partial\omega_{0}/\partial p |_{p=0} = \partial\gamma_{0}/\partial p |_{p=0} = 0$, so let us consider $p=0$. Then Doppler-shifting our expression for $\Omega_{0}(x,p)$ given in \Eqn{equ_2} and differentiating with respect to $x$, we find
\begin{equation}
\left.\frac{\partial \omega_0}{\partial x}\right|_{p=0}=a^{(1)}_r-n\Omprime+2 a^{(2)}_r x
\label{equ_8a}
\end{equation}
and
\begin{equation}
\left.\frac{\partial \gamma_0}{\partial x}\right|_{p=0}=a^{(1)}_i+2 a^{(2)}_i x
\label{equ_8b}
\end{equation}
where $a^{(m)}_{r,i}=\sum_{k}{a^{(m)}_k}$ and the subscripts $r$ and $i$ indicate the real and imaginary component respectively. We require $\partial\gamma_{0}/\partial x =0$, which then
provides an equation for the mode's radial position, $x=x_{0}=-a_{i}^{(1)}/2a_{i}^{(2)}$. Substituting this into $\Eqn{equ_8a}$ we find the critical shearing rate, $\Omprimem$ for which $\partial\omega_{0}/\partial x =0$ at this same value of $x=x_{0}$:-
\begin{equation}
\Omprimem=\frac{1}{n}\left[a^{(1)}_r-\frac{a^{(2)}_r a^{(1)}_i}{a^{(2)}_i}\right]
\label{equ_8c}
\end{equation}
Thus, we expect the maximally unstable isolated mode to exist for this critical shearing rate, $\Omega_{\phi}^{\prime} = \Omega_{\phi m}^{\prime}$, but centred on $x=x_{0}$ rather than $x=0$. Considering the full profile variation case (curve-d), our numerical results show the growth rate peaks at $\Omprime\approx-0.017$ with a value of $\gamma=0.136$, which is $\sim 10\%$ higher than the growth rate for zero flow shear ($\gamma=0.119$). For this case $a^{(1)}=-0.668+0.269i$ and $a^{(2)}=0.361-6.68i$. Substituting these values into \Eqn{equ_8c} provides $\Omprimem=-0.0168$, which is in good agreement with curve-d of \Fig{fig_8}.

We can compare our flow shear results with the global simulations performed in \Ref{P.Hill_1}. This is complicated as we employ a toroidal flow, while the simulations of \Ref{P.Hill_1} employ an 
$\bf{E}\times\bf{B}$ flow, which is almost poloidal. Nevertheless, if parallel flows have a negligible impact, the two can be related by a geometric factor. We can factor out this geometric factor by considering the ratio of the flow shear which maximises the growth rate to the value required to stabilise the mode. Our result of $-0.38$ then agrees very well with that of \Ref{P.Hill_1}, which is $-0.39$.

\Fig{fig_9} presents the scaling of $\Omprimem$ and offset in the poloidal angle
$p_{0}$ with toroidal mode number $n$. As expected from \Eqn{equ_8c} $\Omprimem$ scales with
$1/n$ (or $\rho_{\star}$). Furthermore, we find that at large $n$ $\left|p_{0}\right| = a + bn^{-\alpha}$, where for our profile choice $a=0.041$, $b=1.88$ and $\alpha=0.925$. We have also investigated the effect of asymmetry in the
growth rate spectrum on the reconstructed global mode structure, and this is
illustrated in \Fig{fig_10}. For $\Omprime=0$ the structure is already tilted,
but increasing flow shear in the negative direction acts to re-align the mode
radially and for a critical value of flow shear, $\Omprime = \Omprimem \approx -0.017$, the
effect of the profile variation is completely compensated, allowing an isolated
mode again to form with largest growth rate, $\gamma=$Max$[\gamma_{0}(x,p)]$. Note the mode is radially shifted slightly relative to $x=0$ at this value of $\Omprime$. 
This is consistent with the above analysis. 
Increasing flow shear even further, beyond the critical value, tilts the mode
structure in the opposite direction and lowers its linear growth rate again.
These results, obtained purely from solutions of GS2 and the higher order
theory, are again in good qualitative agreement with global calculations of linear
electrostatic ITG modes presented in \Ref{P.Hill_1}. 

To summarise, for realistic and experimentally relevant cases where we take 
the profile variations into account, we do not in general expect to find pure isolated modes.
Isolated modes can form only in special radial locations where the equilibrium profiles produce
a stationary point in $\Omega_0(x,p)$. However, making adjustments to one equilibrium profile while the others are fixed,
can produce the required stationary point and lead to the onset of the isolated mode, as arises in the above example 
for a critical toroidal flow shear equal to $\Omprimem$. 
~~~~~~~
~~~~~~~
~~~~~~~
\section{Conclusion}\label{conclusion}

In this work we have reconstructed the 2D global mode structure and the global growth rate for linear
electrostatic ITG modes, using local solutions from a gyrokinetic code (GS2) and higher order ballooning theory. This approach, which is solid provided that equilibrium quantities vary slowly across rational surfaces, has provided additional insight into the physics of global simulations of linear microinstabilities in tokamak plasmas.
Our first investigations used radial profiles for the
mode drives that were peaked and symmetric about $x=0$, and we held all other equilibrium profiles
constant; this results in the local complex mode frequency, $\Omega_{0}(x,p)$, having a 
stationary point at $x=0$. This condition produces a special class of mode, known as the 
``isolated mode'', that peaks at the outboard mid-plane with a large growth rate, $\gamma \sim
$Max$[\gamma_{0}(x,p)]$. These results are in very good qualitative agreement
with the simplified fluid model of ITG modes presented in \Ref{Dickinson_1}.
Introducing radial variation into other equilibrium profiles, we show that
the radial position of the stationary points in both local frequency,
$\omega_{0}$ , and growth rate, $\gamma_{0}$, become shifted with respect to
each other. In this case, the reconstructed global mode becomes less unstable
and shifts poloidally away from the outboard mid-plane. 

Toroidal flow shear, introduced as a Doppler shift in the real frequency, also influences the
global mode. Starting from the conditions of an isolated mode, with drive profiles peaked
and symmetric about $x=0$ and with no other profile variations, adding a constant flow shear is always found
to be stabilising. When other profile variations are included, flow shear can be destabilising 
when the flow shear counteracts the tilting of the mode structure at the outboard mid-plane 
that is induced by the other profile variations. This results in an asymmetry in the growth 
rate as a function of flow shear about $\Omprime=0$, which is in qualitative agreement with previous global gyrokinetic
calculations \cite{P.Hill_1}. Moreover, flow shear is also found to shift the mode radially. For a critical flow shear (or a critical toroidal mode number for a given flow shear -- see \Eqn{equ_8c} )
the isolated mode can exist even with arbitrary profiles.

In this paper we have focussed mainly on studying special surfaces 
where fast growing isolated modes arise because of a peaked drive that 
results in a stationary point in $\Omega_{0}(x,p)$. For an arbitrary equilibrium the drive for a particular 
mode number may indeed have local maxima on some special surfaces, which  
can be located using local gyrokinetic solutions to find stationary points in $\Omega_0(x,p)$.
It is this strongly growing isolated mode that will typically be captured by global, initial value gyrokinetic codes. Nevertheless, away from these special surfaces we do not expect to find high growth rate isolated modes, 
but slower growing general modes will predominate. 
(We note that local codes cannot distinguish between isolated and general modes without including higher order corrections.)
This might have important implications for quasilinear transport models, and for flow generation mechanisms, though
other effects associated with nonlinearities and turbulence will also be important: fully nonlinear global 
simulations are needed to assess the relative importance of the linear physics that is described here. 
In this paper we have exploited the higher order analysis to obtain the global mode structure 
and frequency of the stronger isolated mode. Future work will assess the impact of the slower growing general 
modes on transport in other regions of the plasma away from stationary points in $\Omega_0(x,p)$.

Finally, we point out that the procedure used in this work is quite general and can be used to
explore more realistic tokamak equilibria and more complicated instabilities, including the effect of shaping 
(e.g. elongation and triangularity) and electromagnetic modes with kinetic electrons. Future work will extend 
our study to explore these effects.

~~~~~~~
~~~~~~~
~~~~~~~
\section*{Acknowledgement}
The main author is extremely grateful to the Ministry of Higher Education in
Kurdistan region of Iraq for the opportunity and funding they provided to study
for a PhD at University of York. This work has also received funding from the
European Union's Horizon 2020 research and innovation programme under grant
agreement number 633053 and from the RCUK Energy Programme [grant number
EP/I501045]. The views and opinions expressed herein do not necessarily reflect
those of the European Commission. The simulations presented were carried out
using supercomputing resources on HECToR and ARCHER in the UK (provided by the Plasma 
HEC Consortium EPSRC grant number EP/L000237/1), and HELIOS in Japan (funded under 
the Broader Approach collaboration between Euratom and Japan). The authors gratefully 
acknowledge fruitful discussions with Peter Hill on our comparisons with global gyrokinetic simulations. We also acknowledge helpful communications with J W Connor.
~~~~~~~
~~~~~~~
~~~~~~~

~~~~~~~
~~~~~~~
~~~~~~~

\end{document}